\documentclass{elsart}
\usepackage{graphicx}
\begin{document}

\newcommand*{\figany}[4]{
\begin{figure}[#1]
\begin{center}
     #2
   \caption{\label{#4}#3}
\end{center}
\end{figure}
}

\begin{frontmatter}

  \title{How do dynamic heterogeneities evolve in time?} \author[Mainz]{Burkhard
    Doliwa}\ead{doliwa@mpip-mainz.mpg.de},\author[Muenster]{Andreas
    Heuer}\ead{andheuer@uni-muenster.de}
  \address[Mainz]{Max Planck Institute for Polymer Research, Mainz, Germany}
  \address[Muenster]{Department of Physical Chemistry, University of M\"unster,
    Germany}
%
\begin{abstract}
We present simulations of a hard disc system and analyze the time
evolution of the dynamic heterogeneities. We characterize the time
evolution of slow regions and slow particles individually. The
motion of slow clusters turns out to be very restricted, i.e. a
cluster is generated and annihilated in a spatial region four
times the size of its maximum extent. The residual motion of the
cluster can be traced back to subdiffusive motion of the
constituent particles and the process of absorption and loss of
adjacent particles. The subdiffusive dynamics is independent of
how long the particles remain slow. Clusters of fast particles
show an even smaller reach, which seems to be due to their short
life time.
\end{abstract}
\end{frontmatter}
Dedicated to Prof. Hans Sillescu on occasion of his $65^{\rm th}$ birthday.

\section{Introduction}
In the last decade many important pieces of information have been
gained about the microscopic details of the glass formation
process by experiments as well as simulations
\cite{Angell:1995,Ngai:2000,Kob:1999}. One specific aspect deals
with the nature of dynamic heterogeneities \cite{Sillescu:1999}.
It has been realized that dynamic heterogeneities are one of the
key ingredients for understanding the nature of non-exponential
relaxation and the decoupling of translational and rotational
dynamics \cite{fujara:1992}.
Their presence has been uniquely shown by
very different experiments like  NMR \cite{Schmidtrohr:1991},
solvation dynamics \cite{Yang:2001}, and optical
\cite{Ediger:1995} and dielectric dynamic hole burning
\cite{Schiener:1996}. Whereas these methods can be used for bulk
probes, fascinating new experiments have been recently developed
which can be also used for mesoscopic samples
\cite{Russell:2000,Deschenes:2001}.

Most experiments have concentrated on two different aspects of
dynamic heterogeneities. First, what is the exchange time scale on
which slow molecules become fast? Second, on which length scale
are slow (or fast) molecules clustered in the sample? For
temperatures several Kelvin above $T_g$ all experiments clearly
revealed that the exchange time scale is of the order of the
$\alpha$ relaxation. This means that on average after one
relaxation process the dynamics of a slow molecule is
statistically uncorrelated with its initial dynamics. This result
has been mainly obtained for the rotational dynamics of the
molecules. Furthermore the length scale of dynamic heterogeneities
could be directly measured via multidimensional NMR experiments
for which the information about the length scale can be obtained
from spin diffusion. Typical length scales are of the order of 3
nm \cite{Tracht:1998,Tracht:1999}.

In recent years several computer simulations on glass forming
systems have been performed
\cite{Perera:1996,Kob:1997,Heuer:1997,Glotzer:1999,Yamamoto:1998},
in order to elucidate the properties of dynamic heterogeneities
above the mode coupling temperature $T_c$ \cite{Gotze:1989} (and thus still far
above $T_g$). Many details about the relevance of dynamic
heterogeneities, their time scale, and length scale are therefore
known today. For example it could be shown that the length scale
strongly increases with decreasing temperature (above $T_c$) and
that the degree of cooperativity depends on the time scale with a
maximum at a few times the $\alpha$ relaxation time
\cite{Bennemann:1999,Donati:1999,Doliwa:2000}. Furthermore a
strong relation between dynamic heterogeneities and the
non-gaussian parameter has been found \cite{Doliwa:1999}.

To the best of our knowledge no specific information is available
about the time evolution of the dynamic heterogeneities in space.
The goal of this paper is to analyze the time evolution of dynamic
heterogeneities for the translational motion of a simple model
glass former. In what follows we analyze how the slow {\it
particles} as well as the slow {\it regions}, i.e. subsets of
adjacent slow particles, move around. Both aspects are necessary
for a full characterization of dynamic heterogeneities. Note that
also most experiments mentioned above are sensitive to the
properties of the slow molecules. Further, we provide a comparison
with the dynamics of fast clusters. This is a first step to back
up possible scenarios of the time evolution of dynamic
heterogeneities, as indicated e.g. in \cite{Ediger:2001}.

\section{Simulation}

We performed simulations on two- and three-dimensional hard sphere
systems. Here we present the results for the 2D system in order to
optimize visualization. The simulation box contains 9201 particles and
the total simulation time is 5000 $\tau_\alpha$.
We choose a Monte Carlo algorithm for the time evolution of the
system. Our main results are for the packing fraction $\rho = 0.77$. In
order to hamper crystallization we choose a polydispersity of
25\%. Periodic boundary conditions are used.

One technical question deals with the determination of slow
particles as well as slow regions. Choosing the mean square
displacement as a criterion may give misleading results since by
chance fast particles can be back at the origin after some time.
In this respect it has proven useful to monitor the respective
neighborhoods of the individual particles. We define the nearest
neighbors via a modified Voronoi construction, taking into account
the different particle sizes \cite{Gavrilova:1996}.  Then
particles, which are slow on a time scale
$\tau$, can be identified via the criterion that the number of
neighbor changes during this time interval is smaller than some
fixed number $k_{NN}$. Particles leaving or entering a
neighborhood are counted as neighbor changes.  Here we choose
$k_{NN}=5$, which means that particles with an exchange of two
neighbors are considered as slow. This is a sensible choice since
on average, a particle is surrounded by six others.
On this basis we define a particle to be slow on a time scale
$\tau$ at time $t$ if there exists a time interval $t\in[t_0,t_1]$
such that $t_1-t_0=\tau$ and the particle performs less than 5
neighbor changes during this time interval.

Regions of slow particles ({\it slow clusters}) are identified via
a straightforward cluster analysis. Starting from a slow particle
one has to check which other slow particles are connected with
this slow particles via other slow particles, i.e. by a chain of
slow nearest neighbors.

\section{Results}

First we show that the time scale on which nearest neighbor
changes occur is strongly related to the time scale of structural
relaxation. For this purpose we define the function $C_5(t)$ which
counts the fraction of particles with less than 5 neighbor changes
between time 0 and time t. In Fig.1 we show the time-dependence of
$C_5(t)$ for different densities. As known from other observables
the dynamics is dramatically reduced when approaching high
densities.  In the inset we show the decay time of $C_5(t)$ as
defined by the criterion $C_5(\tau_{nn}) = 1/e$, together with the
alpha relaxation time $\tau_\alpha$ which has been extracted from
the incoherent scattering function (see \cite{Doliwa:2000} for more details).
Evidently $\tau_\alpha$ and $\tau_{nn}$ are basically proportional
to each other. Thus $\tau_{nn}$ is also an appropriate time scale
to characterize structural relaxation.
\figany{!ht}{\includegraphics[width=8cm]{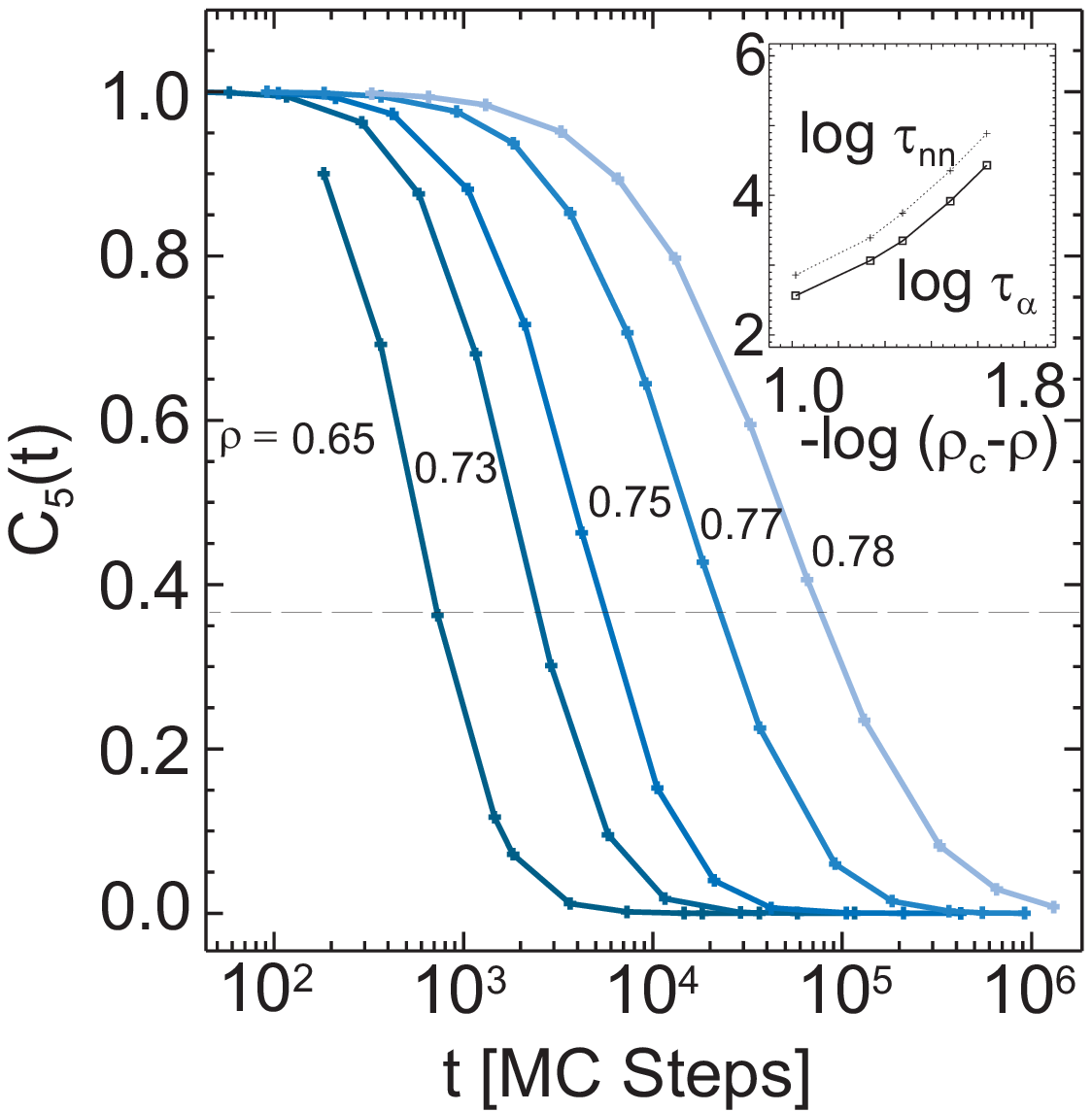}}{
Time dependence of $C_5(t)$ for different densities.
In the inset the density dependence of $\tau_{nn}$ is compared
with that of the alpha relaxation time $\tau_\alpha$.
}{FIG1}
For the analysis of cluster dynamics, we define slow particles on
the time scale $\tau=200\tau_\alpha$. In this way we select ca.
$7\%$ of all particles for the density $\rho=0.77$ which will be
dealt with exclusively now. Before quantifying the time evolution
of slow clusters we visualize the dynamics of a
single
slow cluster,
defined at $t = 0$.
In Fig.2(a) configurations are shown for $t=0$
and $t = 25 \tau_\alpha$. The particles belonging to the
slow cluster are highlighted
by black circles. One can clearly see
that the constituents of the slow cluster as well as some adjacent
particles
on the left side
are moving highly
collectively. The dynamics closely resembles the motion of a small
crystallite in a fluid. Note, however, that the slow cluster is
fully amorphous, showing no crystalline structural features.
Of course, there also exist some other slow particles, which do
not belong to the selected slow cluster.
The dynamics of the same
cluster during the time interval $[50\tau_\alpha,75\tau_\alpha]$,
has many features of the initial dynamics. Again we see the
cooperative dynamics of this system. Comparison of Figs.2(a) and
(b) also shows that the full cluster has shifted
rightwards.
\figany{!ht}{\includegraphics[width=14cm]{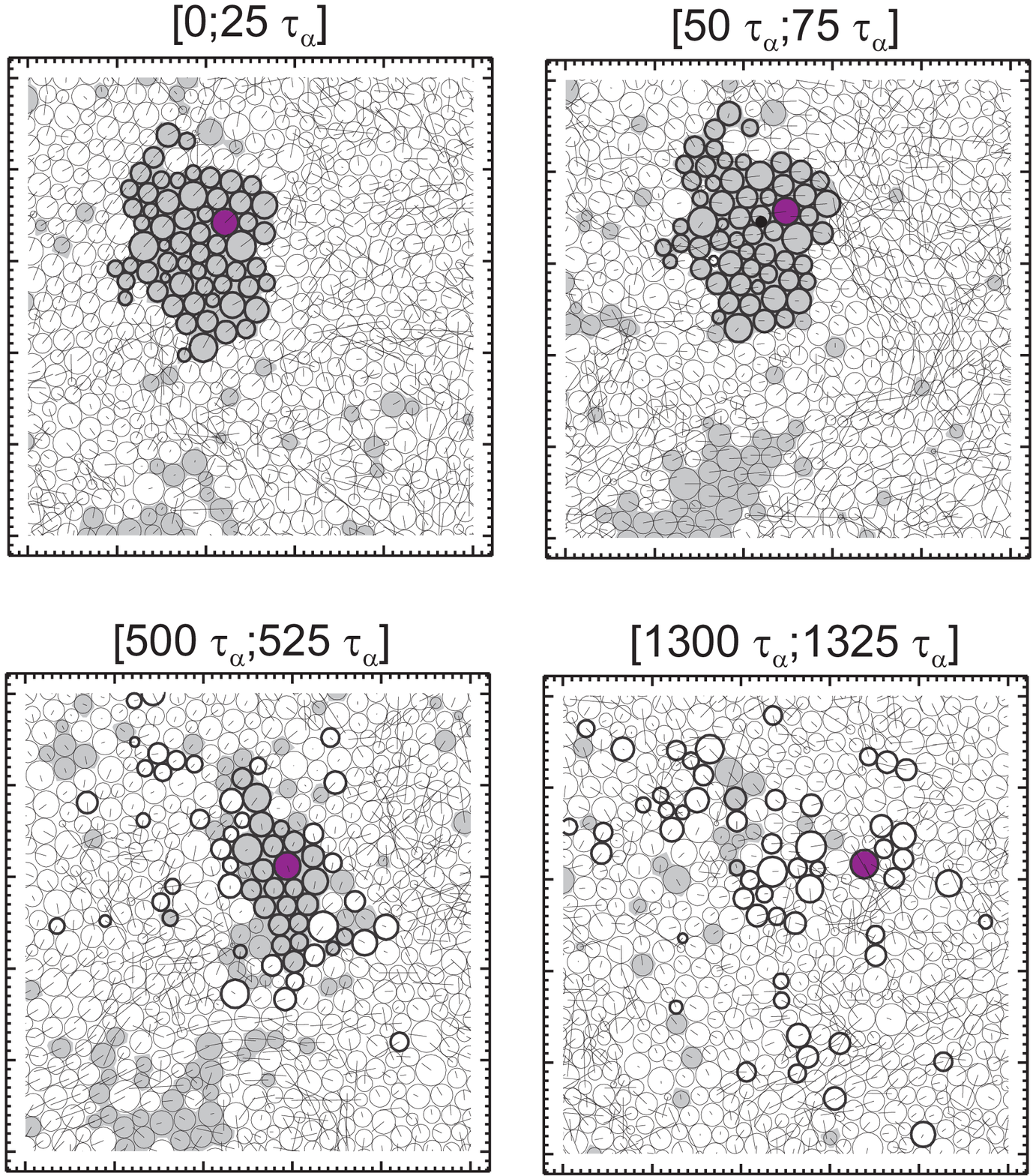}}{
Time evolution of a slow cluster. The constituents of
the slow cluster are highlighted. Shown are the configurations in
the intervals
(a)~$[0,25\tau_\alpha]$,
(b)~$[50\tau_\alpha,75\tau_\alpha]$,
(c)~$[500\tau_\alpha,525\tau_\alpha]$, and
(d)~$[1300\tau_\alpha,1325\tau_\alpha]$.
The configurations at
the beginning and the end of the time interval are connected by
straight lines.
The shaded Voronoi polygons reflect the set of slow particles
at the beginning of the above intervals.
The position in (a) of
the dark particle is shown as a black dot in (b).
}{FIG2}
Thus there exists some type of cluster dynamics, which, however, seems
to be very slow.  In Fig.2(c), showing the dynamics during the time
interval $[500\tau_\alpha,525\tau_\alpha]$ one can clearly see that
several particles from the surface have left the cluster and thus
became fast during $75\tau_\alpha$ and $500\tau_\alpha$. On a
qualitative level this scenario may be denoted {\it surface-melting}.
From the shaded Voronoi areas we see that the cluster also may grow
by attaching new slow particles.
Finally in Fig.2(d) the cluster breaks in its
individual parts - the life time of the slow cluster is over.  Note
that this cluster has lived for a time as long as $1300\tau_\alpha$.

This cluster was one of the longest-living structures in our
simulation. Nevertheless the time evolution is typical for many
other slow clusters appearing and disappearing during the course
of the simulation. In particular we always observe the highly
cooperative nature of the dynamics. This observation directly
implies that it is not possible for particles to enter into the
core of a slow cluster. We have not seen such an event during our
simulations. The dynamics of the cluster has two facets. First,
the highlighted particles were moving
also during the time where
they belonged to the slow cluster. We call this {\it real}
dynamics.
Additionally, the set of particles
belonging to the slow cluster, may vary with time.
From Fig.2(b) to (c),
the slow cluster has lost 38 members but gained 17 new.
This type of time evolution will be
denoted as {\it fictive} dynamics.

In order to quantify the time evolution of dynamic heterogeneities
we will be guided by two questions: How does the slow cluster move
in space during its life time and which particles contribute to
this slow cluster?
To answer these questions,
we introduce four observables: (i) A is the absolute area one
distinct slow cluster covers during its lifetime. For the
definition of A the Voronoi areas of all members of the cluster at
all times are superimposed. (ii) M is the maximum cluster size
during its life time. We only consider clusters with $M > 10$
average particle sizes.
(iii) $A_p$ is defined in analogy to $A$
but counts the total number of different particles which were
members of the cluster during its life time. (iv) $M_p$ denotes in
analogy to $M$ the maximum size of the slow cluster in terms of
particles. In particular we are interested in the ratios $A/M \ge
1$ and $A_p/M_p \ge 1$.  First we discuss the possible scenarios
in terms of both values.
If dynamic heterogeneities were dominated
by extrinsic forces (e.g. dynamics of interacting ions in a
basically fixed disordered network) there would exist fixed areas in
space where the dynamics would be slower all the time. Whenever a
particle entered this area it would become slow. Since many particles
will enter and leave one expects $A/M \ll A_p/M_p$.
Intuitively this means that the (here totally immobile) slow regions
move slower than the slow particles.
In contrast, if the dynamic
heterogeneities are not related to fixed structural variations in
the probe it is hard to imagine that the dynamics of slow regions
is slower than the dynamics of the constituting slow particles.
This general physical reason implies $A/M \ge A_p/M_p \ge 1$.
In the case $A/M \approx 1$ the time evolution of the slow cluster
can be characterized as a local process of generation and annihilation.
No dynamics is
involved. For $A/M> 1$ the dynamics of the slow regions can be characterized by two
independent contributions which
correspond to the real and fictive
dynamics, introduced above.
First, in the case $A_p/M_p = 1$ the
dynamics of the slow region is fully related to the dynamics of
the constituting slow particles, thus it is real. Second, in the
opposite case $A/M = A_p/M_p$ the dynamics of the slow cluster is
exclusively due to changes in the set of slow particles belonging
to the slow cluster. Thus it is fictive.

\figany{!ht}{\includegraphics[width=10cm]{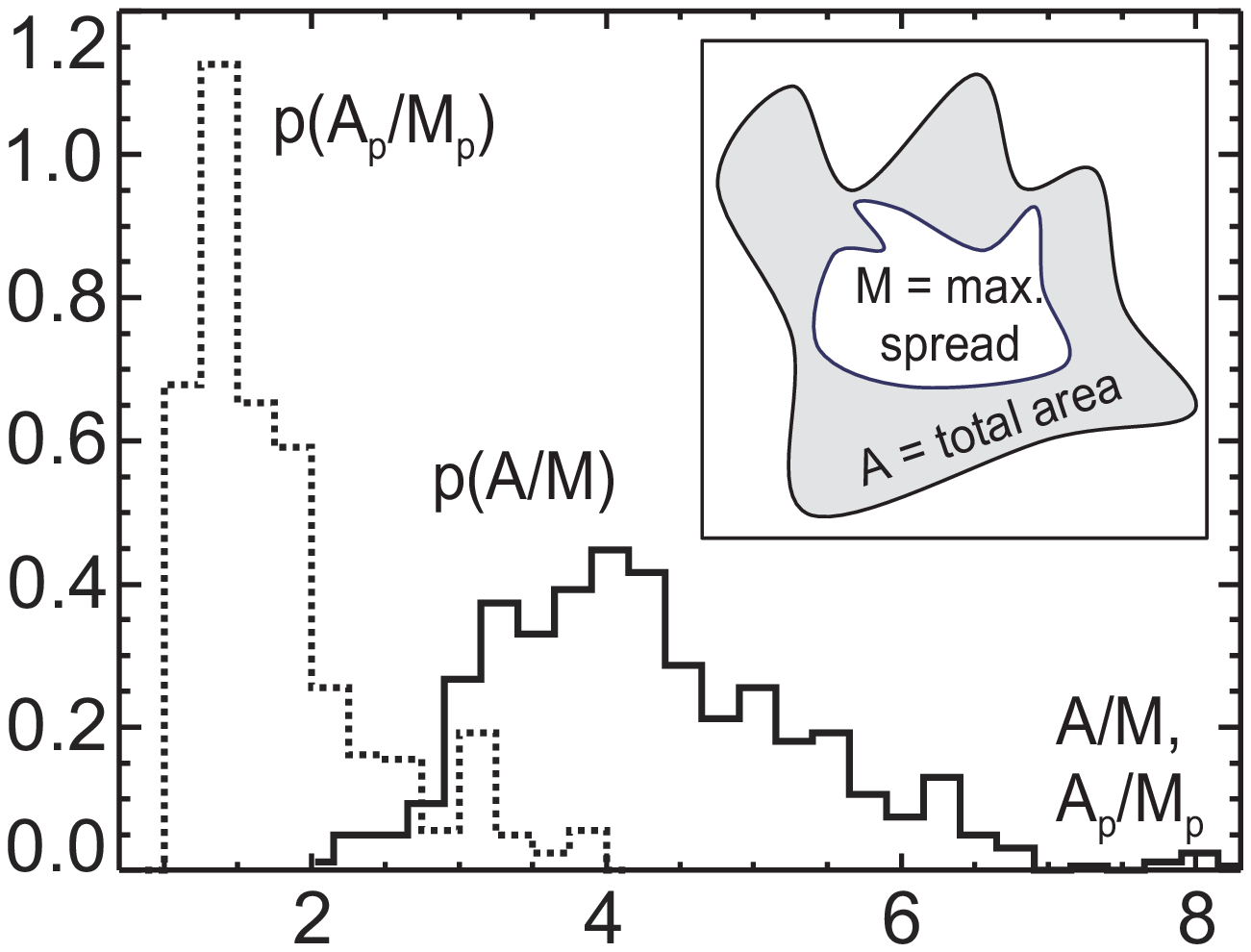}}{
The distribution of the ratio $A/M$ and $A_p/M_p$
(see text) for all relevant slow clusters.
The mean values are $4.3$ and $1.8$, respectively. The inset
illustrates the definition of $A$ and $M$.
}{FIG3}
For our analysis we only consider clusters with $M > 10$ average
particle sizes.
In Fig.3 the distribution of $A/M$ and $A_p/M_p$
is plotted for all relevant clusters during our analysis. The
average value of $A/M$ is $4.3$. This result indicates that in
agreement with the example of Fig.2 there is
only a minor
time evolution of the slow clusters.
However,
the generation and annihilation process
of a typical slow cluster is not fully local. As
discussed above, the value of $A_p/M_p$ contains
the
relevant information
about the number of particles involved in
the time evolution of the clusters.
One can
see that its distribution is
shifted to
smaller values as compared to $A/M$. We obtain $\langle A_p/M_p
\rangle = 1.8$. A cluster of maximum size ten rambles therefore
(on average) between 18 particles. The ratio
$(\langle A_p/M_p\rangle-1)/(\langle A/M \rangle-1)$ is a good measure for the
contribution of fictive dynamics to the overall dynamics.
Here we see that approximately $3/4$ of the cluster dynamics is due
to real and one quarter due to fictive dynamics.
Using a shorter time scale for the identification of slow particles (here we used
$\tau=200\tau_\alpha$, see above) one would
increase the contribution of fictive dynamics since
it would be more easy for the slow cluster to attach adjacent particles.
If we eventually reached a selection level of more than ca.~15\%, the life time
of clusters would become infinite due to fictive dynamics. In other words,
there would always remain fragments of every cluster which would further
bequest their origin to freshly emerging slow regions.
For this reason, we have restricted the calculation of $A/M$ to the 7\% slowest
particles, corresponding to a selection time of $\tau=200\tau_\alpha$.

\figany{!ht}{\includegraphics[width=10cm]{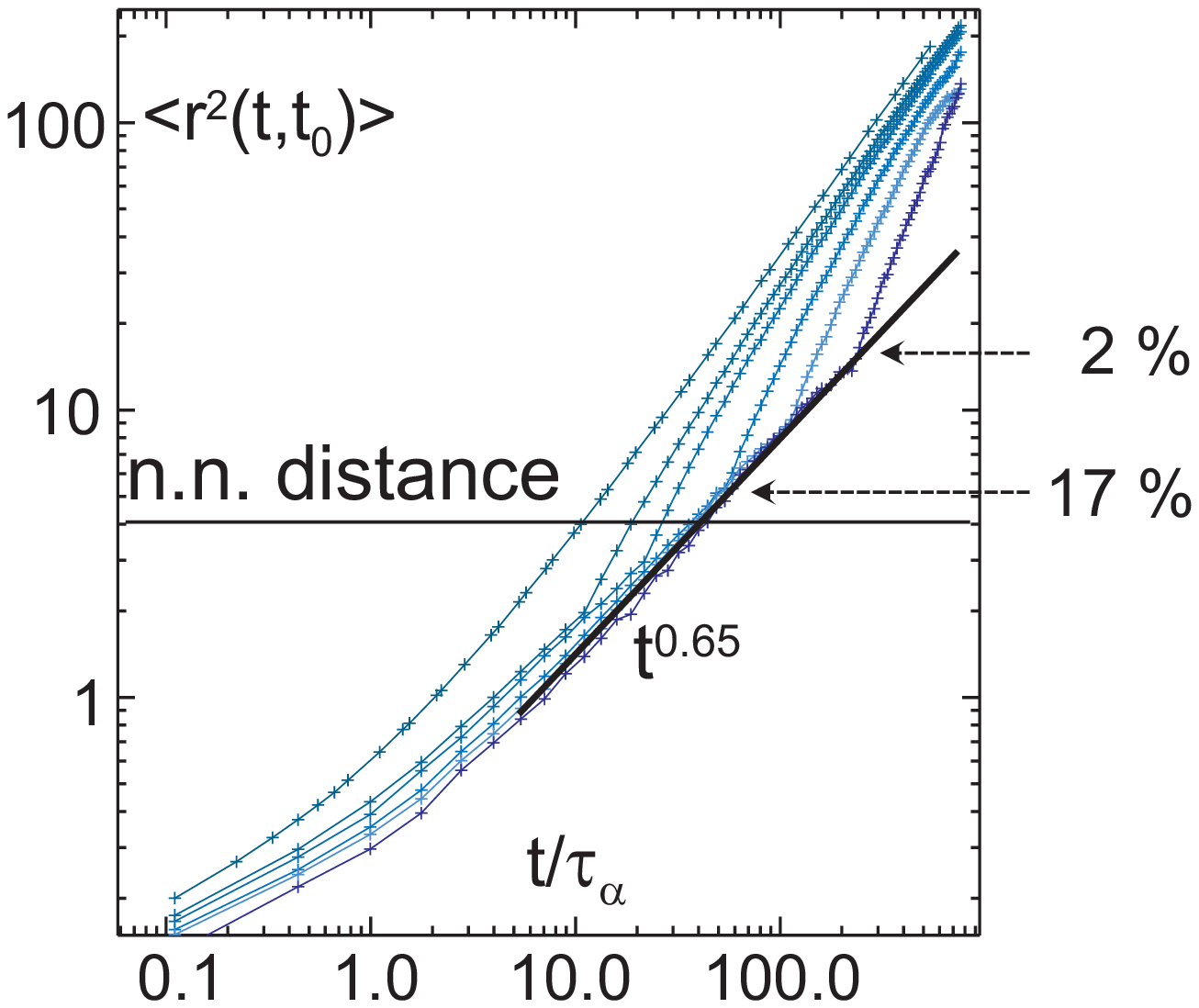}}{
The mean-square displacement
$\langle r^2(t,t_0)\rangle$ for particles which
are slow during $t=0$ and $t=t_0$ and
have their fifth neighbor change immediately afterwards.
The different
lines correspond to (from the second left to right) $t_0 =
10,20,50,100,200 \tau_\alpha$.
The left curve
corresponds to the mean-square displacement, averaged over all
particles.
For $t_0=50\tau_\alpha$ and $t_0=200\tau_\alpha$,
the fractions of particles with less than 5 neighbor changes
are 17\% and 2\%, as can be extracted from $C_5(t)$ of Fig.1.
}{FIG4}
We have seen that also the sluggish dynamics of the constituting
particles is a relevant effect for the dynamics of slow clusters.
In any event, most experiments are only sensitive to the single
particle behavior. Therefore it may be of interest to study the
single particle dynamics somewhat closer. Here we consider the
mean square displacement $\langle r^2(t,t_0)\rangle$ of
particles which
have less than 5 neighbor changes during the time
interval $[0,t_0]$, and
have the 5th neighbor change during the subsequent interval  $[t_0,t_0+\Delta t]$,
where $\Delta t=t_0/10$.
We first discuss the case $t_0=200\tau_\alpha$
which corresponds to the rightmost curve in Fig.4.
One can clearly see
the change of slope around $t=t_0$. This effect is directly
related to the fact that by definition all particles become mobile
around $t_0$. Of interest is the time regime $t<t_0$ during which
the particles were slow.
The dominant feature here is the subdiffusive
dynamics with an exponent 0.65 during two decades of time. This
quantifies the previous observation in Fig.2 that slow particles
(and thus slow regions) are not totally immobile. Intuitively,
the subdiffusive behavior implies that slow clusters are rattling
back- and forth in effective cages formed by the surrounding fluid
like particles. Interestingly, the curves for the other values of
$t_0$ look very similar for their respective $t < t_0$. Thus the
life time of a slow cluster and the dynamics during this life time
are uncorrelated. From the numbers extracted from $C_5(t)$ (see
Fig.1, one can estimate that more than 80\% of all particles
become fast before moving more than a nearest neighbour distance,
i.e. $\langle r^2 \rangle \approx 4$. Thus on the length scale of
the nearest neighbor distance a slow particle can be viewed as
immobile until it becomes fast and subsequently contributes to the
structural relaxation.

So far, we have concentrated on the regions of slow particles. One
might expect that the time evolution of fast regions is quite
different. To check this, we defined fast particles by not being
slow on a time scale $\tau=10\tau_\alpha$, which again yielded a
selection of about $7\%$. Here we obtain $\langle A/M\rangle=3.4$
and $\langle A_p/M_p \rangle=2.8$. Thus, fast clusters are more
localized than slow clusters.
Furthermore, the ratio
$(\langle A_p/M_p\rangle-1)/(\langle A/M\rangle-1)=0.75$
indicates that the real dynamics has only a little contribution to the
overall time evolution. This may come as a surprise since we are
concerned with very mobile particles. Obvious explanations for this are
the comparatively small life time of fast clusters and the lacking
cooperativity of fast particles.

\section{Discussion}
Using appropriately defined observables, computer simulations can
be used to obtain very detailed information about microscopic
processes. Here we have specified the time evolution of slow
particles and slow clusters. As main results we obtained that
(i)~the dynamics of particles in slow clusters is highly
cooperative, thus excluding the dynamics of single particles
through slow clusters, (ii)~the dynamics of slow clusters is
highly restricted in space, (iii)~the residual dynamics is both
due to temporal variation in shape as a result of adjacent fast
particles becoming slow or slow particles from the surface
becoming fast (fictive dynamics) and due to the motion of the
constituting particles (real dynamics), (iv)~the so-defined real
dynamics is subdiffusive until the particle becomes fast and is
independent from the time the particle remains slow, (v)~clusters
of fast particles move less than their slow counterparts. On a
qualitative level the results (i)-(iii) can be already identified
in Fig.2. They might give rise to some refinement of previous
models of dynamic heterogeneities, see e.g. in
\cite{Cicerone:1997,Diezemann:1997}.

Finally we would like to discuss two questions. What is the
relation to experimental results and what is the underlying origin
of these features which are at the core of the glass-forming
process?

A discussion of the relation to experiments has to contain the
statement that the time scales of experiments and simulations are
vastly different. Therefore it is important to analyze whether or
not properties of the glass formation process around $T_g$ and
$T_c$ are different. Furthermore most experimental results have
been gained for rotational dynamics whereas most simulations have
concentrated on translational dynamics. It has been shown,
however, from simulations of a molecular glass former that already
above the mode coupling temperature the dynamic heterogeneities of
rotational and translational dynamics are coupled in a nearly
maximum way. Molecules with fast translational dynamics also
rotate fast and vice versa \cite{Chen:1997,Qian:1999}.
In the
temperature range several Kelvin above $T_g$ experiments have
revealed that on the time scale of $\tau_\alpha$ a slow molecule
becomes fast \cite{Ediger:1995,Heuer:1995,Sillescu:1996}.
Note that during $\tau_\alpha$ the mean square
displacement of a molecule is of the order of the squared nearest neighbour
distance. Therefore the experimental result, generalized to translational
dynamics, implies that typical slow molecules become fast after moving a
nearest neighbour distance. This is indeed the case, see Fig.4.
According to measurements by the Ediger group \cite{Wang:1999}
the behavior
changes at $T_g$ where the life time of the slow molecules
becomes much larger than the structural relaxation time.

The question of the origin of dynamic heterogeneities is very
complex. One answer has been found in terms of energy landscapes
for a Lennard-Jones system. For very small systems $N = 60$
particles it has been shown that the slow regions correspond to
configurations very deep in the potential energy landscape
\cite{Buchner:2000}. High activation barriers have to be passed so
that the system can become mobile again. Some correlations with
the local potential energy have been also observed in simulations
of larger Lennard-Jones systems \cite{Donati:1999}. For the hard
disc system the potential energy landscape is not a convenient way
of representation of different configurations. Rather one would
like to use concepts like locally stable structures. We have seen
that obvious quantities like particle sizes or local density fail
to identify locally slow regions. It seems that the underlying
reason for regions of the system to be slow is to be found in
complex multi-particle correlations. These correlations prevent a
local region to relax since, in analogy to a crystal, internal
degrees of freedom are not present. The observation that the
dynamics of a cluster is independent of its life time and thus of
its stability shows that this is mainly related to the interaction
with the adjacent fast particles and not with the internal
structure. Currently we are trying to identify the relevant
structural quantities which are the basis of dynamic
heterogeneities.

We gratefully acknowledge helpful discussions with H.W. Spiess.
This work has been supported by the DFG, Sonderforschungsbereich 262.

\bibliographystyle{unsrt}
\bibliography{Refs}

\end{document}